\documentclass[10pt,a4paper]{article}
\usepackage{lineno}
\usepackage{nameref}
\usepackage{graphicx}
\usepackage{cite}
\usepackage{authblk}
\usepackage{color}
\usepackage{amsmath}
\usepackage{graphicx}
\usepackage{array}
\usepackage{caption}
\usepackage{subcaption}
\usepackage{geometry}
\begin{document}

\title{The Escaramujo Project: instrumentation courses during a road trip across the Americas \\
\vspace{0.5cm}
\normalsize
Proceeding of the 38th International Conference on High Energy Physics\\
		3-10 August 2016\\
		Chicago, USA }
		
\author{Federico Izraelevitch \thanks{fogo@fnal.gov, Present Address: Instituto Dan Beninson, Universidad Nacional de San Mart\'{i}n-Comisi\'{o}n Nacional de Energ\'{i}a At\'{o}mica, Argentina }\\
Fermi National Accelerator Laboratory, Batavia, IL, United States of America.}
        
        \date{}
\maketitle

\abstract{The Escaramujo Project was a series of eight hands-on laboratory courses on High Energy Physics and Astroparticle Instrumentation, in Latinamerican Institutions. The Physicist Federico Izraelevitch traveled on a van with his wife and dogs from Chicago to Buenos Aires teaching the courses. The sessions took place at Institutions in Mexico, Guatemala, Costa Rica, Colombia, Ecuador, Peru and Bolivia at an advanced undergraduate and graduate level. During these workshops, each group built a modern cosmic ray detector based on plastic scintillator and silicon photomultipliers, designed specifically for this project. After the courses, a functional detector remained at each institution to be used by the faculty to facilitate the training of future students and to support and enable local research activities. The five-days workshops covered topics such as elementary particle and cosmic ray Physics, radiation detection and instrumentation, low-level light sensing with solid state devices, front-end analog electronics and object-oriented data analysis (C++ and ROOT). Throughout this initiative, about a hundred of talented and highly motivated young students were reached. With the detector as a common thread, they were able to understand the designing principles and the underlying Physics involved in it, build the device, start it up, characterize it, take data and analyze it, mimicking the stages of a real elementary particle Physics experiment. Besides the aims to awaken vocations in science, technology and engineering, The Escaramujo Project was an effort to strengthen the integration of Latinamerican academic institutions into the international scientific community.}

\section{Introduction and project summary}
\normalsize
The instrumentation schools organized in Latinamerican countries over the past decades (e.g. the ICFA instrumentation school series) have shown an increasing interest in Particle Physics detection technologies in the region. The strengthening of the participation of groups in experiments at first-class international facilities (such as CERN, Fermilab, Pierre Auger Observatory, etc.) and a thriving cohort of young Latinamerican experimentalists in the field is a clear accomplishment of those efforts. Detector and instrumentation schools can be a pivotal experience in the career of a student. Hence, aligned with these endeavors, we organized and lectured a series of them in the way of a drive from Chicago to Buenos Aires: the Escaramujo Project\footnote{www.escaramujo.net}. 

\begin{figure}[ht]
  \begin{center}
    \includegraphics[width=0.5\textwidth]{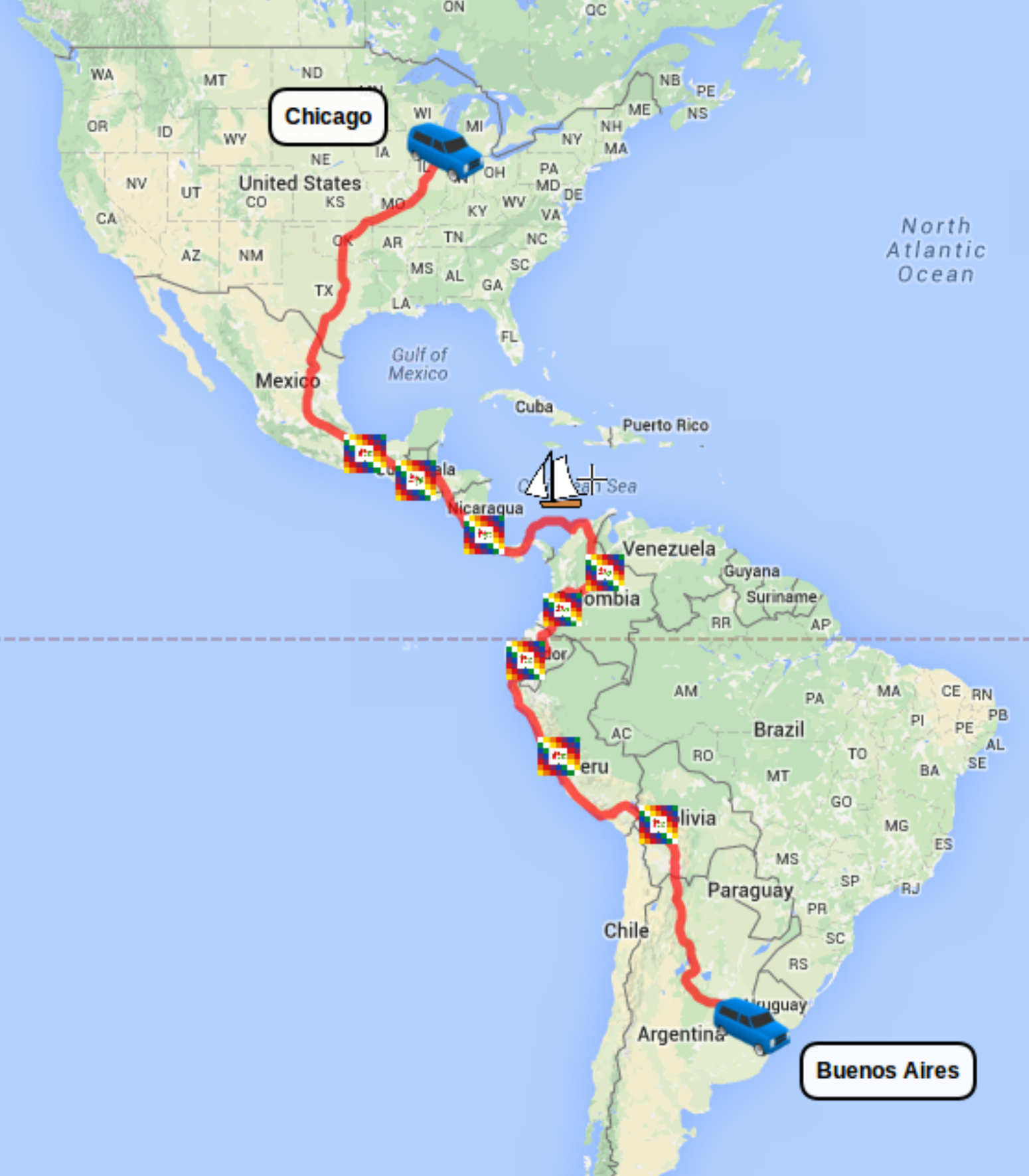}	\caption{Map of the courses and the route followed.}	\label{fig:Chicago-BuenosAires_new}
  \end{center}
\end{figure}

\begin{figure}[ht]
  \begin{center}
    \includegraphics[width=0.7\textwidth]{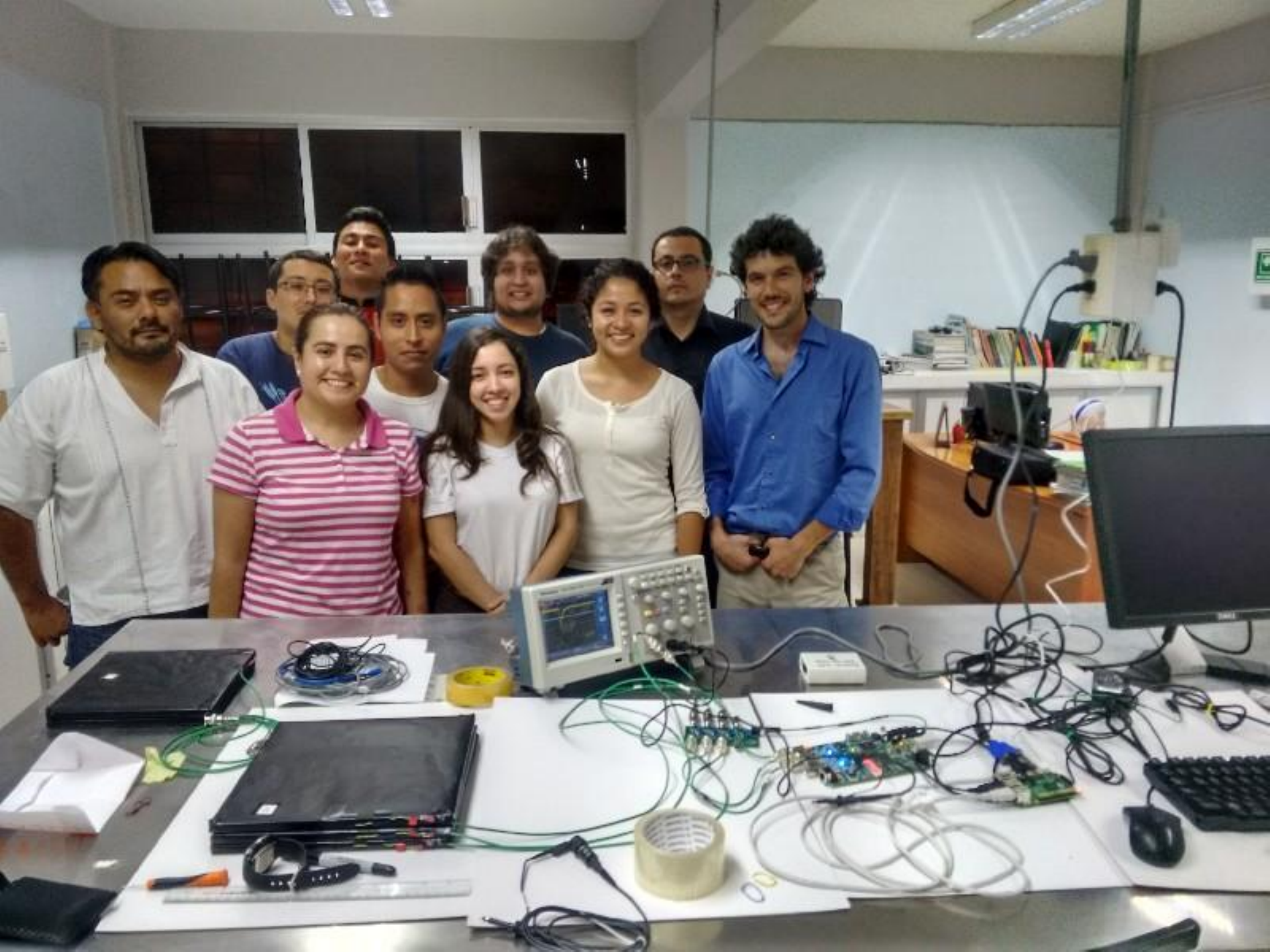}	\caption{Participants of The Escaramujo course at the Universidad Aut\'{o}noma de Chiapas, M\'{e}xico, with the detector in the forefront.}	\label{fig:setupUNACH}
  \end{center}
\end{figure}

We completed eight hands-on instrumentation laboratory courses at Latinamerican Institutions along our way. The sessions were attended by advanced-undergraduate and graduate students of Science, Technology, Engineering and Mathematics careers. As a common thread to explore Particle Physics Instrumentation, we designed a modern, simple and cost-effective cosmic rays detector, based on plastic scintillator and silicon photomultiplier (SiPMs). At each site, groups of students assembled the detector from a set of parts provided by the project. After each course, a functional detector remained at the visited institution as a donation to be used by local professors to facilitate the training of future students and to support and enable local research activities. The course sites were chosen with a draft route in mind, based on existing relations with professors and scientists and prioritizing already members of the Latinamerican Giant Observatory (LAGO). At each site, a local organizer was responsible of the students selection and the preparation of the classrooms, laboratories and a minimum of equipment (oscilloscope, gloves, etc.). A map showing the visited institutions and the routed followed can be seen in figure \ref{fig:Chicago-BuenosAires_new}. Figure \ref{fig:setupUNACH} is a picture of the participants of the Escarmujo course at the Universidad Aut\'{o}noma de Chiapas, Mexico. Table \ref{tab:sites} compiles the information of the visited institutions, courses dates and local organizers.

\begin{table}[ht] 
\begin{center}
  \begin{scriptsize}
  	 \begin{tabular}{ m{6 cm} l l m{4 cm}}
     \hline \hline
     Institution & Country & Date & Local Organizer	\\
     \hline \hline
     Universidad Aut\'{o}noma de Chiapas \mbox(UNACH)  and Mesoamerican Centre for Theoretical Physics (MCTP)  & Mexico & Aug-2015 & Karen Caballero and Arnulfo Zepeda \\ \hline
     Universidad de San Carlos (USAC) & Guatemala & Sep-2015 & H\'{e}ctor P\'{e}rez and Edgar Cifuentes \\ \hline
     Universidad de Costa Rica (UCR) & Costa Rica & Oct-2015 & \'{A}lvaro Chavarr\'{i}a (KICP-UC) and Arturo Ram\'{i}rez \\ \hline
     Universidad Industrial de Santander (UIS) & Colombia & Nov-2015 & Luis N\'{u}\~{n}ez  \\ \hline
     Aut\'{o}noma Universidad de Nari\~{n}o \mbox(AUNAR) and Universidad de Nari\~{n}o (UdeNar) & Colombia & Dec-2015 & Jaime Betancourt \\ \hline
     Universidad San Francisco de Quito (USFQ) and Escuela Polit\'{e}cnica Nacional (EPN) & Ecuador & Dec-2015 & Dennis Cazar, Edgar Carrera and Nicol\'{a}s V\'{a}zquez \\ \hline
     Comisi\'{o}n Nacional de Investigaci\'{o}n y Desarrollo Aeroespacial (CONIDA) & Per\'{u} & Jan-2016 & Luis Otiniano \\ \hline
     Universidad Mayor de San Andr\'{e}s (UMSA) & Bolivia & Jan-2016 & Hugo Rivera and Mart\'{i}n Subieta \\      
	 \hline \hline
	 \end{tabular}
	 \caption{Visited institutions, courses dates and local organizers names.}
	 \label{tab:sites}
  \end{scriptsize}
\end{center}
\end{table}

\section{The detector}

\begin{figure}[ht]
  \begin{center}
    \includegraphics[width=0.8\textwidth]{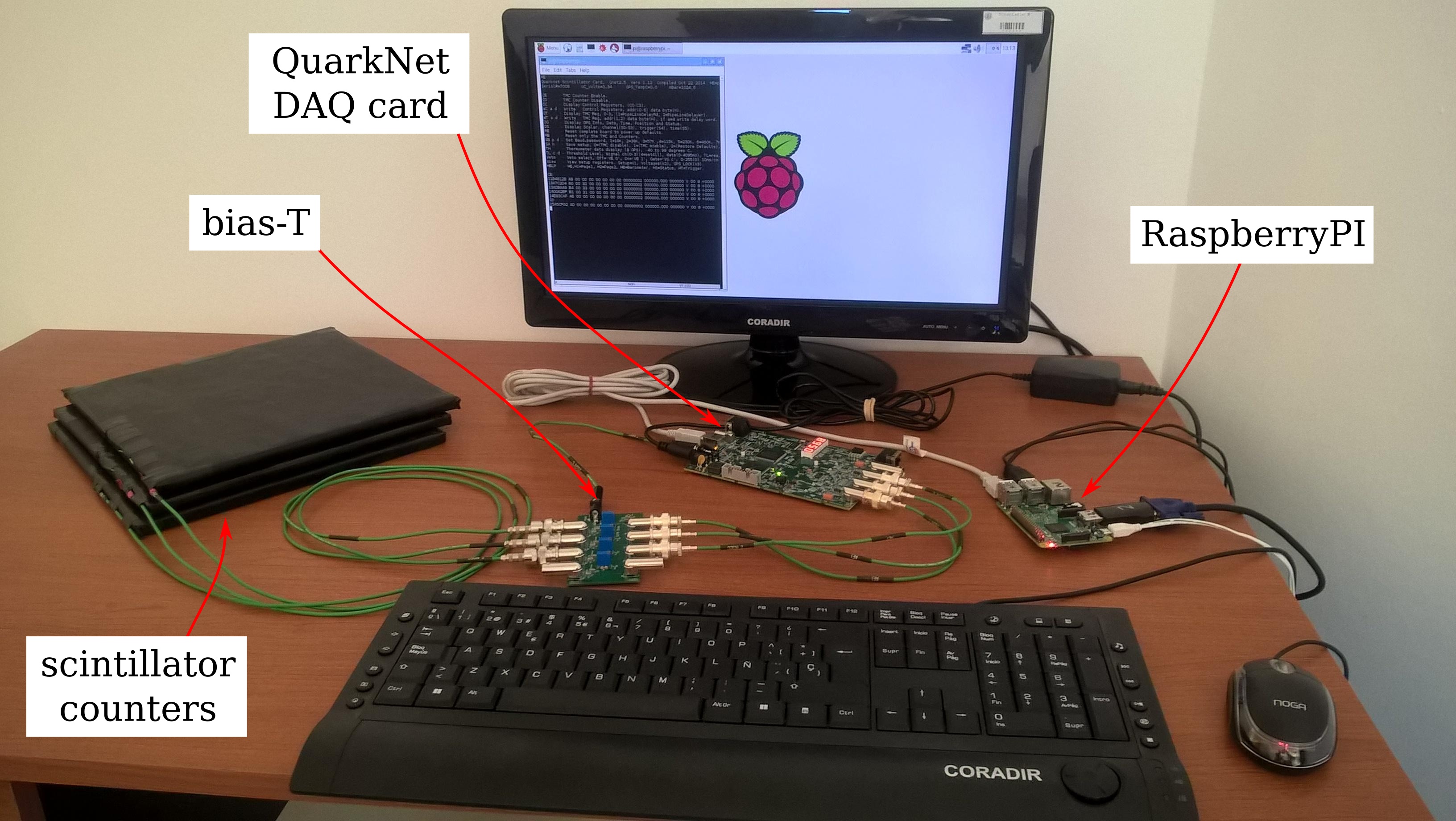}	\caption{The Escaramujo detector setup.}	\label{fig:fullSetup}
  \end{center}
\end{figure}

The design of the detector and its scope were a trade-off of many factors. First of all, we wanted the device to be assembled at each course by students. It was desirable that as many tasks as possible were completed by the attendees. However, considering that one of the objectives was that the detector were functional by the end of the course, we sought to minimize the risks at troubleshooting. We then prepared and tested all the electronic boards, cables and connectors before shipping. Secondly, we aimed to bring state-of-the-art detection technology to the visited institutions. Thirdly, because of logistics reasons compactness was crucial. Finally, simplicity and cost-effectiveness were overarching principles for the design. The setup, shown in figure \ref{fig:fullSetup}, consisted of plastic scintillators coupled to SiPMs, readout by a data acquisition board connected to a minicomputer. 

\subsection{Scintillator and photodetector}

\begin{figure}[ht]
  \centering
  \begin{subfigure}[b]{.5\textwidth}
    \centering
    \includegraphics[height=0.8\textwidth]{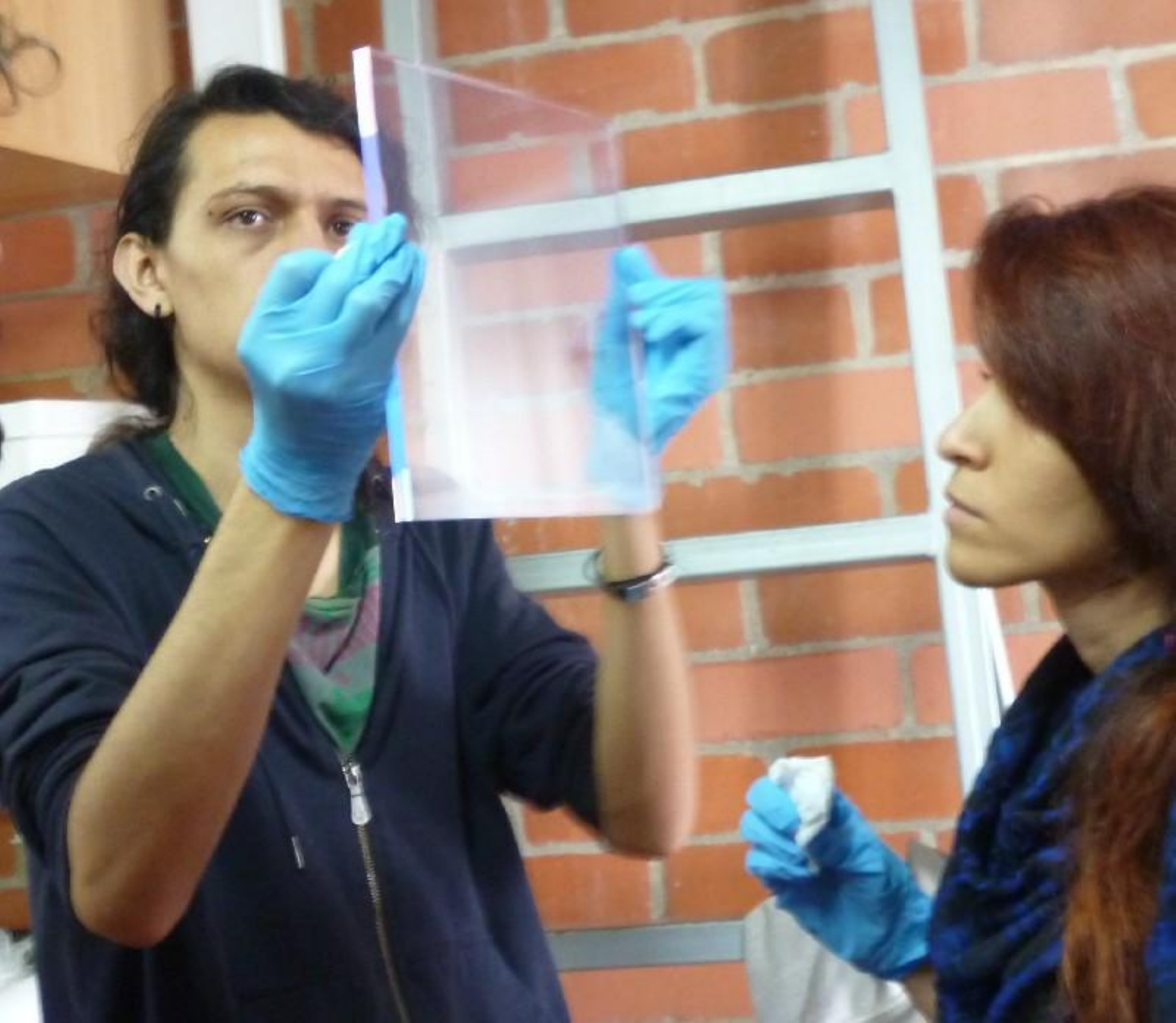}	
  \end{subfigure}%
  \begin{subfigure}[b]{.5\textwidth}
    \centering
    \includegraphics[height=0.8\textwidth]{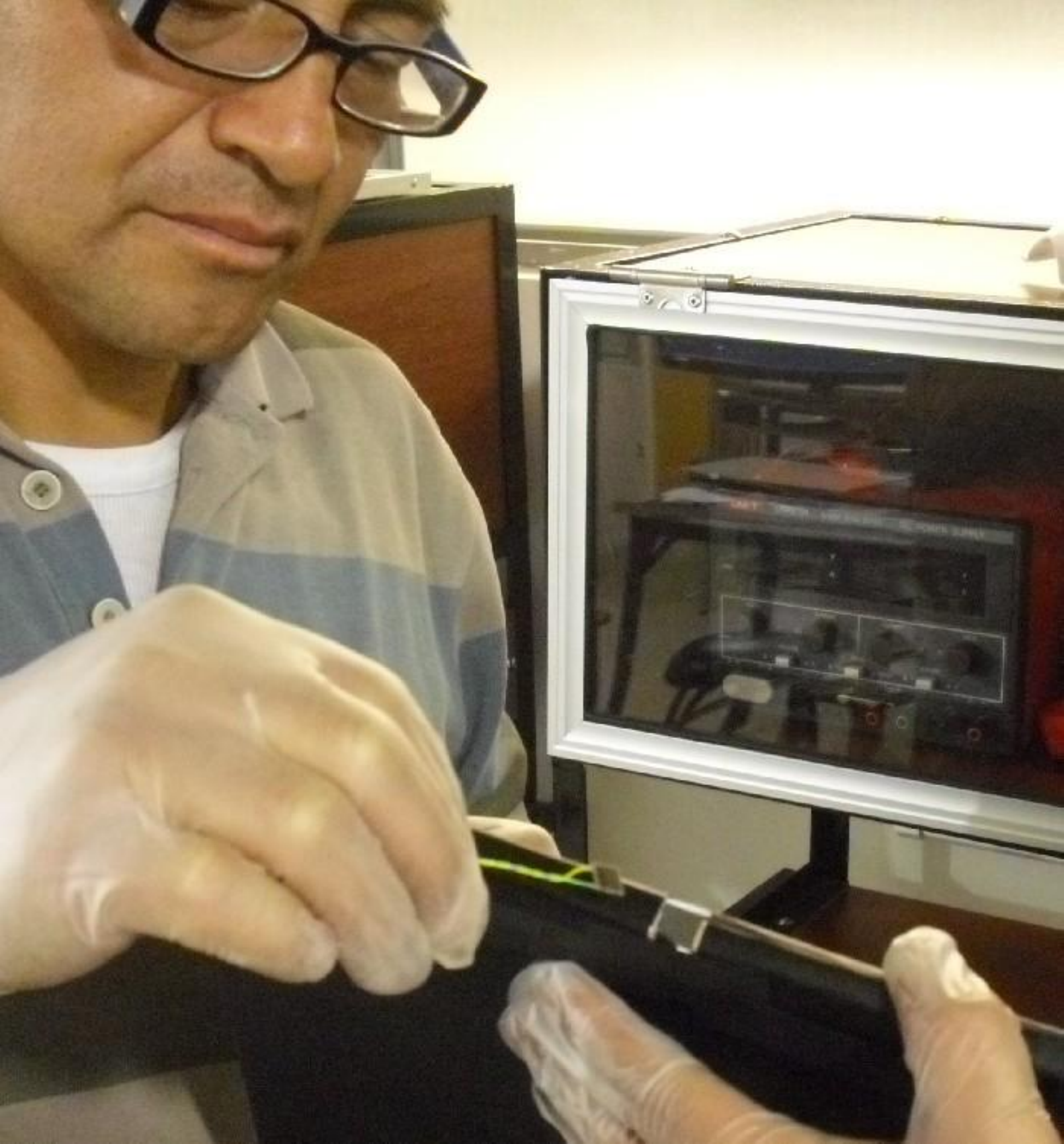}	
  \end{subfigure}
  \caption{Left: two Physics student cleaning a scintillator plate before wrapping. Right: Professor Jaime Betancourt (Universidad de Nari\~{n}o, Colombia) coupling the SiPM to the scintillator.}	
  \label{fig:scint_SiPM}
\end{figure}

The detector was comprised of three plates of scintillator (EJ-200, Eljen Technology) coupled to SiPMs (MicroFC-60035-SMT, SensL). The size of the plates was 25~x~25~x~1~cm$^3$. Each plate was readout by one SiPM, coupled to one of the sides of the plate with the gentle pressure of adhesive tape. We used neither a coupling material nor optical fibers. The coupling was made as simple as the SiPM propped against the scintillator. The counter was wrapped with an inner layer of Tyvek paper and an outer layer of black opaque paper. A hole was cut in both layers of paper as a feedthrough, to allow connection with the SiPMs. The wrapping was sealed with black-electrical tape to ensure light tightness. Figure \ref{fig:scint_SiPM} shows a bare scintillator plate and one SiPM being coupled to the side of a wrapped one. 

\subsection{Data acquisition}

The data acquisition (DAQ) was based on the QuarkNet DAQ card version 6000 \cite{cite:quarkNet}. It consists of four channels, each having an amplifier (voltage amplitude gain of 10) and an edge discriminator with software-adjustable threshold in updating configuration (i.e. the output pulse stays high as long as the input signal level is above the threshold). The digital part of the board is essentially a time-to-digital converter (TDC) with a complex programmable logic device (CPLD). The TDC records the arrival time of the leading and trailing edges of the discriminator output, with 1.25~ns resolution. The CPLD allows the user to define the trigger logic (1- to 4-fold coincidence), record scalers of single channels and coincidence counts, define the discriminator thresholds, window gates, among other functions. When a trigger is fired, the system generates an event that contains the times of all the edges within the window gate referenced to a global timestamp. The DAQ card is powered with 5~V~DC and has a 5~V fan-out connector to bias other devices. 

The QuarkNet DAQ card version 6000 was conceived to work with photomultiplier tubes. To adapt it for the use with SiPMs, we designed a four-channels \textit{bias-T} board. The board takes the 5~V from the fan-out connector of the QuarkNet card and generates $\approx$~30~V with the LT3571 DC-DC converter. A coupling capacitor prevents the DC from flowing to the QuarkNet card input. The bias voltage of each SiPM can be adjusted with a trimming potentiometer, while it is being monitored on a test point.

The readout of the QuarkNet DAQ card can be conveniently done via USB standard. The setup was completed with low-cost single-board computer (Raspberry Pi 2 Model B), on which we installed Minicom, a text-based serial-port communication program used to talk to the QuarkNet card. Through native QuarkNet commands, it was possible to configure the board and start and stop the data acquisition. The raw data was stored in a plain text file and then copied to a PC for data analysis.

\section{Courses organization}

The courses lasted 4 or 5 days, accommodating  the academic schedule of the visited institution. They included chalk board lectures and slides presentations, laboratory sessions and data analysis classes, usually in a computer's lab. The idea that governed the courses was that the students could plow through all the stages of a real Particle Physics experiment: a) the understanding of the Physics phenomenon to investigate and the conception of a setup to do it, b) the comprehension of the designing principles of the detector, c) its fabrication, d) the characterization and the determination of the operating parameters, e) data taken and f) analysis.

The lectures covered topics in elementary particle Physics, passage of radiation through matter, low-level light sensing, DAQ electronics and object-oriented data analysis (C++ and ROOT). The visited institutions had a large diversity of experience and expertise in the mentioned topics. Therefore, the lectures were tailored to accommodate the background of the students according to their curriculum. Besides the participants of the hands-on laboratory sessions, the lectures were usually attended by a larger audience comprised of other interested students.

In the laboratory sessions each of the three scintillator counters was assembled by a group of two or three students (some of the times professors participated in the assembly, too). Laboratory tasks were guided with on-the-fly explanations. The course participants cleaned and wrapped the scintillator plates with both layers of paper and coupled it to the SiPM. Each group was responsible for the light tightness of the counters. The cable connections and the start up of the detector were also done by the students. To characterize the detector and decide the operating parameters of it (SiPM bias voltages, discriminator thresholds, window gates, etc.), the participants discussed and performed several experiments. For example, with the sandwich configuration (the top and bottom counters as trigger and the middle one as the one under study) the threshold and bias voltage of each counter was varied and optimized. After the characterization, Physics experiments like muon-flux angular dependence and muon decay-time measurement were done. Finally, depending on the available time and local interests, some extra activities were explored. For example, in La Paz, Bolivia, students installed the detector in a minivan and measured the cosmic ray flux as a function of the altitude while driving from La Paz (3650 m a.s.l.) to the Chacaltaya Observatory (5240 m a.s.l.). In Pasto, Colombia, course participants built an ad hoc support for a muon telescope. Students and professors had a field day to Galeras volcano to measure its muon-flux attenuation using the Escararmujo detector. 

The courses included computer sessions, in which students learned to write basic ROOT scripts to analyze the data taken by themselves. We developed a C++/ROOT unpacker that, given a QuarkNet raw-data text file, produces a TTree with the information of each event. This unpacker was given to the students, along with an outline of an analysis script. The deepness of the analysis part of the courses strongly depended on participants background in programming. 

\section{Outlook}

Throughout this initiative, about a hundred of talented and highly motivated young students were directly reached. Several sites have been organizing their own outreach and training sessions using the donated detector, enlarging the impact of the Escaramujo Project. As all the detectors are identical, most of the groups are starting to interact to each other to share and analyze the data they collect, organizing a collaboration. There are universities that were not visited that are building their own detector based on the Escaramujo design. There are also plans to organize more courses in the near term, to continue the task of awakening motivations in experimental particle Physics.

\section{Acknowledgements}

Two companies provided the necessary components to complete the detector kits for the courses. SensL\footnote{www.sensl.com} donated the silicon photomultipliers and Eljen Technology\footnote{www.eljentechnology.com} the plastic scintillators. Fermilab contributed with the donation of the DAQ and readout for all the setups. The Kavli Institute for Cosmological Physics through the Research Associate \'{A}lvaro Chavarr\'{i}a, and the Universidad de Costa Rica provided extra funds that allowed a larger laboratory course in Costa Rica (with four detectors and the invited lecturers Pablo Mosteiro and \'{A}lvaro Chavarr\'{i}a --also the course local organizer). All the host institutions provided lodging and meal expenses for the visiting teacher and family while staying for the courses. Institutional support was provided by the Argentinian Science and Technology Ministry through its program RAICES (Spanish acronym of Network of Argentinian Researches and Scientists working Abroad), the Consulate General of Argentina in Chicago with the support of its Consul General Marcelo Su\'{a}rez Salvia, the Argentinian Scientists Network in the Midwest of the U.S.A. (RECARMO, from its Spanish acronym) with its coordinator Mario Zaritzky and the LAGO coordinator Hern\'{a}n Asorey. Thanks to Otto \'{A}lvarez, Marcela Carena, Juan Estrada, Sten Hansen, Alina Lusebrink, Jorge Montes, Erik Ramberg, Paul Rubinov and many others that supported the project by different means. Special thanks are due to Andrew Lathrop for his support and encouragement.

\end{document}